\documentclass[onecolumn,showkeys,a4paper,twoside]{revtex4-1}

\usepackage{color}
\usepackage{latexsym}
\usepackage{amsmath}
\usepackage{amssymb}
\usepackage{graphicx}% Include figure files
\usepackage{dcolumn}% Align table columns on decimal point

\usepackage{subcaption}
\usepackage{tcolorbox}
\usepackage{hyperref}
\hypersetup{
    colorlinks,
    citecolor=blue,
    filecolor=black,
    linkcolor=black,
    urlcolor=blue
}
\usepackage[french,english]{babel}
\usepackage[latin1,applemac]{inputenc}
\usepackage[T1]{fontenc}
\usepackage{times}

\input epsf

\usepackage{empheq}
\definecolor{myblue}{rgb}{0.53,0.8,1}
\newcommand*\mybluebox[1]{%
\fcolorbox{black}{myblue}{\hspace{1em}#1\hspace{1em}}}

\begin{document}

\selectlanguage{\english}

\title{On the concept of a generalized law of refraction: A phenomenological model \\
The Article and the Supporting Informations}
\author{Emmanuel Rousseau and Didier Felbacq}
\affiliation{Laboratoire Charles Coulomb, UMR5221 CNRS-Universit\'e de Montpellier, 34095, Montpellier, France}
\vspace{1cm}
%\date{\today}

\begin{abstract}
This paper presents investigations on the generalized laws of refraction and reflection for metasurfaces made of diffractive elements. It introduces a phenomenological model that reproduces all the features of the experiments dedicated to the generalized Snell-Descartes laws. Our main finding is that the generalized laws of refraction and reflection as previously stated have to be modified in order to describe the propagation of light through metasurfaces made of diffractive elements. We provide the appropriate laws that take a different form depending on the properties of the metasurface. Our models apply to both periodic and non-periodic metasurfaces. We show that the generalized law of refraction strictly exists only for linear-phase profiles and sawtooth-wave phase profiles under constraints that we specify. It can be approximatively defined for non-linear phase profiles.\\
This document includes the article as the part \ref{P.1} and the supporting informations as the part  \ref{P.2}.
\end{abstract} 


\keywords{Nanophotonics, Metasurfaces, Snell-Descartes Law, Diffractive Optics, Refraction}

\maketitle

\tableofcontents

\newpage

\part{\textit{The Article}: \\ On the concept of a generalized law of refraction: A phenomenological model}
\label{P.1}

\vspace{1cm}
\textbf{Abstract:}
This paper presents investigations on the generalized laws of refraction and reflection for metasurfaces made of diffractive elements. It introduces a phenomenological model that reproduces all the features of the experiments dedicated to the generalized Snell-Descartes laws. Our main finding is that the generalized laws of refraction and reflection as previously stated have to be modified in order to describe the propagation of light through metasurfaces made of diffractive elements. We provide the appropriate laws that take a different form depending on the properties of the metasurface. Our models apply to both periodic and non-periodic metasurfaces. We show that the generalized law of refraction strictly exists only for linear-phase profiles and sawtooth-wave phase profiles under constraints that we specify. It can be approximatively defined for non-linear phase profiles.

\section{Introduction: Snell-Descartes laws and their generalization}

\begin{figure}[h]
   \begin{center}
   \includegraphics[width=10cm]{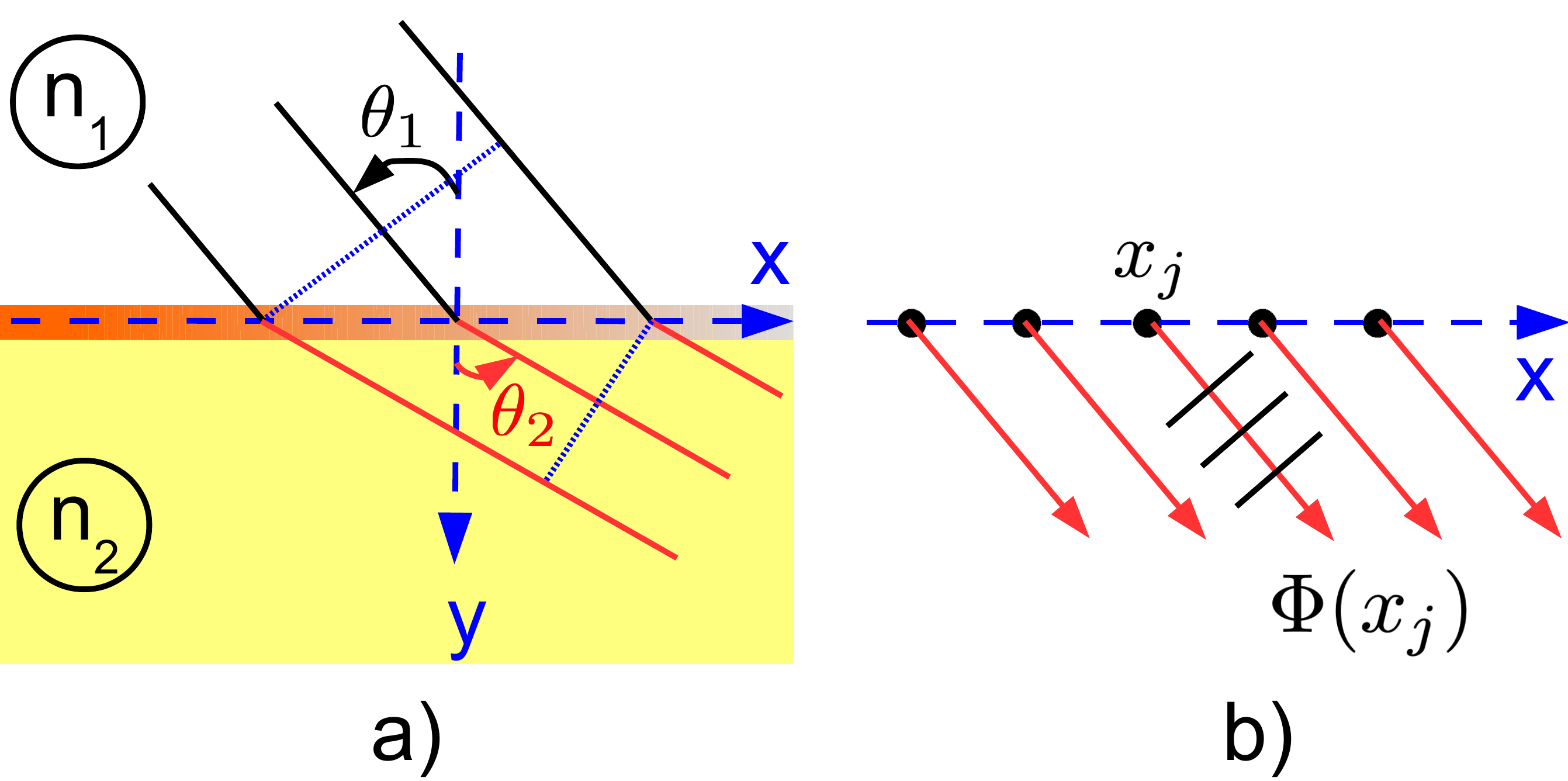}
   \end{center}
   \caption
   { \label{fig1} a) Metasurface contribution to refraction. An incident plane-wave from a medium with refractive index $n_1$  is refracted to a medium with refractive index $n_2$. The metasurface is sketched by a gradient color from red to gray. The angle of incidence is $\theta_1$ whereas the refracted angle is $\theta_2$. The black and red lines represents respectively the wave-vector direction of the plane-waves in the medium of incidence and in the medium of transmission.   b) Metasurface made by a collection of diffractive elements (scatterers). Each dots represents a scatterer that radiates a plane wave with an extra phase-term $\Phi(x_j)$ as compared to the incident plane wave, $x_j$ being the scatterer position.}
\end{figure}

Metafilms\cite{Kuester_2003} and metasurfaces\cite{Holloway_2012,Yu_2011} have been introduced as layers  that are thin at the wavelength scale. In the spirit of metamaterials studies where the contribution of the heterogeneities can be averaged and an effective homogeneous behavior can be defined\cite{Felbacq_2016}, metasurfaces have been shown to give an extra contribution  to the reflection and refraction of plane-waves \cite{Yu_2011,Ni_2011}. As a consequence, generalized laws of reflection and refraction have been considered in order to take into account the contribution of metasurfaces to light propagation\cite{Aieta_2012b,Yu_2011,Larouche_2012} . Considering the refraction of a plane wave between two media of refractive indexes $n_1$ and $n_2$ separated by a metasurface [see Fig.(\ref{fig1}-a)], the generalized law reads \cite{Yu_2011}:
\begin{align} 
\label{law}
n_2 \sin \theta_2- n_1\sin \theta_1 = \frac{1}{k_0}\frac{d\Phi(x)}{dx}
\end{align} 
where $\theta_1$ is the angle of incidence, $\theta_2$ is the angle of refraction, $k_0$ the wave-vector in vacuum, $\Phi(x)$ is an extra phase term that the metasurface adds to the total phase of the refracted plane wave. The phase function $\Phi(x)$ may vary along the interface according to the $x$-variable. It is sometime referred to as a phase-discontinuity\cite{Aieta_2012b,Ni_2011,Yu_2011} or a phase-jump\cite{Larouche_2012}.

The quantity $D(\theta_1,\theta_2) = n_1 \sin \theta_1- n_2\sin \theta_2$ appears in different model of optics. The Snell-Descartes law of refraction\cite[p.125]{Born_1985} reads $D(\theta_1,\theta_2) = 0$. It is the cornerstone of the geometrical optics model where $\theta_1$ (\textit{resp.} $\theta_2$) is the angle between the incident (\textit{resp.} refracted) light ray and the normal to the interface. The relation $D(\theta_1,\theta_2) = 0$ is also met when dealing with the transmission of a plane wave between two media with respective refractive indices $n_1$ and $n_2$ \cite[p.38]{Born_1985}. In such a case, the angles $\theta_1$ and $\theta_2$ are the angles between the wave-vector of the incident and transmitted plane waves and the normal to the interface. Finally, the quantity $D(\theta_1,\theta_2)$ is also relevant in diffractive optics where it enters the grating equation  (\cite[p.499]{Hecht_2015} and  \cite[p.22]{Palmer_2002}) that reads $D(\theta_1,\theta_2) = m \frac{2\pi}{k_0 \Lambda}$ where $m \in \mathbb{Z}$ is the order of the principal intensity maxima, $k_0$ the wave-vector in vacuum and $\Lambda$ the period of the grating. The grating equation specifies the directions represented by the angles  $\theta_1$ and $\theta_2$ respectively to the normal of the grating for which the interferences are constructive, leading to maxima of the light intensity, the so-called diffraction orders. If $\frac{k_0}{2\pi} \Lambda < 1$, only $m=0$ can solve the grating equation. In such a case, the grating equation formally simplifies to the Snell-Descartes law of refraction.

Metasurfaces have been said to lead to a generalization of the Snell-Descartes law \cite{Aieta_2012b,Yu_2011}. Given that it appears in different context as seen above, one can 
 wonder for which of the previously mentioned situations that generalization is relevant. In ray optics, the quantity $\frac{d\Phi}{dx}$ appearing in eq. (\ref{law}) is meaningful and is a local one: it corresponds to the derivative of the phase discontinuity at the position $(x,y=0)$ where the ray of light crosses the interface between the two media located at y=0. On another hand, the $x$ dependence of the quantity $\frac{d\Phi(x)}{dx}$ makes it ambiguous when considering the refraction of a plane-wave, for there is no  "x" position that can be chosen, since the plane wave extends all over the interface. Moreover, for a non-linear phase function $\Phi(x)$, the transmitted wave may be of complicated shape, \textit{i.e.} not a plane wave. An example of such a situation is considered in the ref.\cite{Ruphuy_2014,Ruphuy_2015} where a phase profile similar to a parabola leads to the focusing of a plane-wave normally incident to the surface.  In such a case, the very notion of angle of refraction loses its meaning. Furthermore, the ref. \cite{Yu_2011}  derived the eq.(\ref{law}) based on the Fermat principle. Such a starting point implies the framework of geometrical optics. On another hand, the generalized law of refraction eq.(\ref{law}) involves a length scale $1/k_0$, proportional to the wavelength of light, hence specific to wave optics. This length-scale is absent from geometrical optics. Finally, a quantity called the phase profile $\Phi(x)$ enters the generalized law of refraction \cite{Yu_2011} whereas the phase is a notion specific to harmonic functions and plane-waves\cite[p.26]{Hecht_2015}. The generalized law of refraction appearing in Yu \textit{et al.}\cite{Yu_2011} mixes concepts from geometrical optics and wave-optics but it is used to describe experiments involving diffractive elements, while the Fermat principle does not take diffraction into account. Consequently, the range of validity and application of the generalized law of refraction has to be clarified and made precise. This is the goal of the present work. The focus here is on the case of diffractive optics where the metasurface is made on a collection of scatterers. This geometry was chosen since this is the way metasurfaces are experimentally implemented in the infrared or visible ranges \cite{Aieta_2012b,Ni_2011,Yu_2011}. It will be shown that a generalized law of refraction can be obtained exactly only in the case of linear and sawtooth phase-profiles. Moreover, within these assumptions, the generalization of the law of refraction will be shown to exist but with a form different from the previously reported results \cite{Aieta_2012b,Yu_2011}. Finally in the case of sawtooth profiles, the generalized Snell-Descartes law will prove to be nothing else than the grating relation.

\section{A phenomenological model}
 
Experimentally, metasurfaces are made of a periodic array of metallic\cite{Huang_2012,Li_2015,Ni_2011,Sun_2012,Yu_2011} or dielectric nanoparticles\cite{Lin_2014,Zhou_2017,Sun_2017,Aoni_2019} (see \cite{Genevet_2017} for a review). Each scatterer can be seen as a secondary light source that radiates a spherical wave. For a scatterer at position $(x_j,0,0)$ on the metasurface, the radiated wave at the observation point in a medium with refractive index $n_2$ reads $\frac{1}{n_2 k_0 |R|} e^{i n_2 \vec{k}_0.\vec{R}}e^{-i \omega t}$ where $\vec{R}=(x-x_j,y,z)$  is the distance between the observation point $(x,y,z)$ and the scatterer; $\vec{k}_0=(k_x,k_y,k_z)$ is the wave-vector in vacuum. 

Let us assume that only observation points located close to the $y$-axis and far from the metasurface are considered [see Fig.(\ref{fig1}-a)]. Under these hypotheses, distances are large as compared to the wavelength in medium $n_2$, \textit{i.e.} $n_2 k_0 |R|\gg 1$ and the paraxial approximation $(x,z) \ll y$ holds. As a consequence, the spherical wave can be approximated locally by its tangent plane. More details are given in the section I of the Supplementary Informations.

Within these assumptions, the electromagnetic field radiated by the nanoparticles can be described by a plane wave  $A_0 e^{i n_2 \vec{k}_0.\vec{R}}e^{-i \omega t}$ \cite[chp.30]{Feynman_2011}. In order to implement the phase discontinuities, the field radiated by the nanoparticles has to be delayed as compared to the incident plane wave\cite{Genevet_2017}. The delay produces a phase difference between the radiated plane and the incident plane wave. Engineering the particle geometry leads to delays that are then particle-dependent. 

Now without loss of generality, but for convenience, we consider a 2D-problem and consider only the plane with coordinate $z=0$. The particle numbered $j$ at position $(x_j,0)$ on the metasurface adds a contribution $\Phi(x_j)$ [see Fig.(\ref{fig1}-b)] to the phase of the radiated plane-wave. The resulting plane wave radiated by this particle is $A_0 e^{i \vec{k}_2.\vec{r}}e^{i\Phi(x_j)}$ where $\vec{k}_2$ is the wave-vector in medium 2. The array of nanoparticles implements the metasurface. The scattered-field distribution is given by the sum of all the plane waves radiated by the nanoparticles. Assuming that $N$ particles constitute the metasurface, the electric field scattered by the metasurface at the observation point $\vec{r} = (x,y)$ reads:
\begin{align} \label{eq:DisSum} 
E(x,y)= A_0 e^{i(\vec{k}_2.\vec{r}-\omega t)} \sum_{j=0}^{N-1} e^{i k_0 D(\theta_1,\theta_2)x_j} e^{i \Phi(x_j)}
\end{align} 
This equation is derived in the supplementary materials as the eq.(S.1). The total electric-field $E_{tot}(x,y)$ in the medium of refractive index $n_2$ is the sum of the field that is scattered by all the scatterers constituting the metasurface and the refraction of the incident plane wave into the medium with refractive index $n_2$ in the absence of the metasurface. The field scattered by the metasurface is $E(x,y)$ as given by the eq.(\ref{eq:DisSum}). The field transmitted in the medium of refractive index $n_2$ is $t(\theta_1,\theta_2) A_{inc} e^{i \vec{k}_2.\vec{r}}$ where  $t(\theta_1,\theta_2)$ is the Fresnel coefficient for transmission. The total electric-field then reads as $E_{tot}(x,y) = E(x,y) + t(\theta_1,\theta_2) A_{inc} e^{i \vec{k}_2.\vec{r}}$. As a consequence, as observed in all the experiments\cite{Huang_2012,Li_2015,Ni_2011,Sun_2012,Yu_2011}, there is always one beam that follows the usual Snell-Descartes law [the contribution $t(\theta_1,\theta_2)A_{inc} e^{i \vec{k}_2.\vec{r}}$] and another one that follows a generalized Snell-Descartes law [the contribution $E(x,y)$]. In the following we consider only the latter. 
The list of the scatterers positions $\{ x_j\}_{j=0,..,N-1}$ can lead to a regular or a random sampling of the phase function $\Phi(x)$. Most of the experimental realizations \cite{Aieta_2012b,Ni_2011,Yu_2011} and propositions consider a regular sampling. Nevertheless a random sampling can also be considered and the hypothesis of regular or random sampling does not limit the present investigation. 

In the limit of a large number of scatterers, the discrete sum eq.(\ref{eq:DisSum}) converge to the continuous sum:
\begin{align} 
\label{eq:ConSum} 
E(x,y)=  \tilde{A}_0 e^{i(\vec{k}_2.\vec{r}-\omega t)} \int_{0}^1 e^{i k_0 L D(\theta_1,\theta_2)\bar{x}} e^{i \Phi(L\bar{x})} d\bar{x} ,
\end{align} 
where $ \tilde{A}_0=NA_0$ is the renormalized amplitude of the electric field scattered by the metasurface and $L$ the spatial extension of the metasurface. The convergence of the discrete sum eq.(\ref{eq:DisSum}) to the continuous sum eq.(\ref{eq:ConSum}) is guaranteed in a regular-sampling scheme because the eq.(\ref{eq:DisSum}) is nothing else than a Riemann sum.The difference between the discrete sum and the continuous sum scales as $\frac{1}{N}$. In a random-sampling scheme, the Monte-Carlo theorem \cite[p.77]{Kalos_2009} ensures the convergence of the discrete sum to the continuous result with a difference that scales as $1/\sqrt{N}$. We can now consider different shapes for the function $\Phi(x)$. We assume that it can be written as a power series $\Phi(x) = \sum_{k=1}^{+\infty} \phi_k x^k$. The constant term is not considered since it can be factorized out in the eq.(\ref{eq:DisSum}-\ref{eq:ConSum})  and then does not contribute to the intensity scattered by the N particles. For a linear function ($\phi_k = 0 ~~\forall k>1$), the light-intensity distribution at infinity is given in the supplementary materials as the equation eq.(S.2) for the discrete sum and as eq.(S.8) in the case of the continuum approximation. 

A generalized Snell-Descartes can now be defined if some relations are satisfied. These relations are eq.(S.5-S.7) for discrete metasurfaces and eq.(S.10) in the continuum limit. They allow for a single plane-wave to emerge from the metasurface with a refractive angle $\theta_2$ given by the eq.(\ref{eq:GenDesc}):
\begin{align}
\label{eq:GenDesc}
n_2 \sin \theta_2 - n_1 \sin \theta_1 & = \frac{1}{k_0} \phi_1 =  \frac{1}{k_0}  \frac{\Phi(x)}{x} .
\end{align}

 The generalized refraction law is similar for both the discrete and the continuum cases. Let us stress that the right-hand term of the generalized Snell-Descartes law eq.(\ref{eq:GenDesc}) is $ \frac{1}{k_0}  \frac{\Phi(x)}{x}$ and not $ \frac{1}{k_0}  \frac{d \Phi(x)}{d x}$ as found in ref.\cite{Aieta_2012b,Yu_2011}. This is due to the fact that, in order to compute the light-intensity distribution, the following quantity $(k_0 D(\theta_1,\theta_2)\bar{x} + \phi_1 \bar{x})$ should be factorized as $(D(\theta_1,\theta_2) + \frac{\phi_1 \bar{x}}{k_0 \bar{x}} )k_0 \bar{x} $. Consequently, the calculation leads to a right-hand term that is $ \frac{1}{k_0}  \frac{\Phi(x)}{x}$. The outcome of our model with its assumptions very close to the experimental realizations, is not the right-hand term $ \frac{1}{k_0}  \frac{d \Phi(x)}{d x}$ found in the ref.\cite{Aieta_2012b,Yu_2011}.
 
\begin{figure}
   \begin{center}
   \includegraphics[width=12cm]{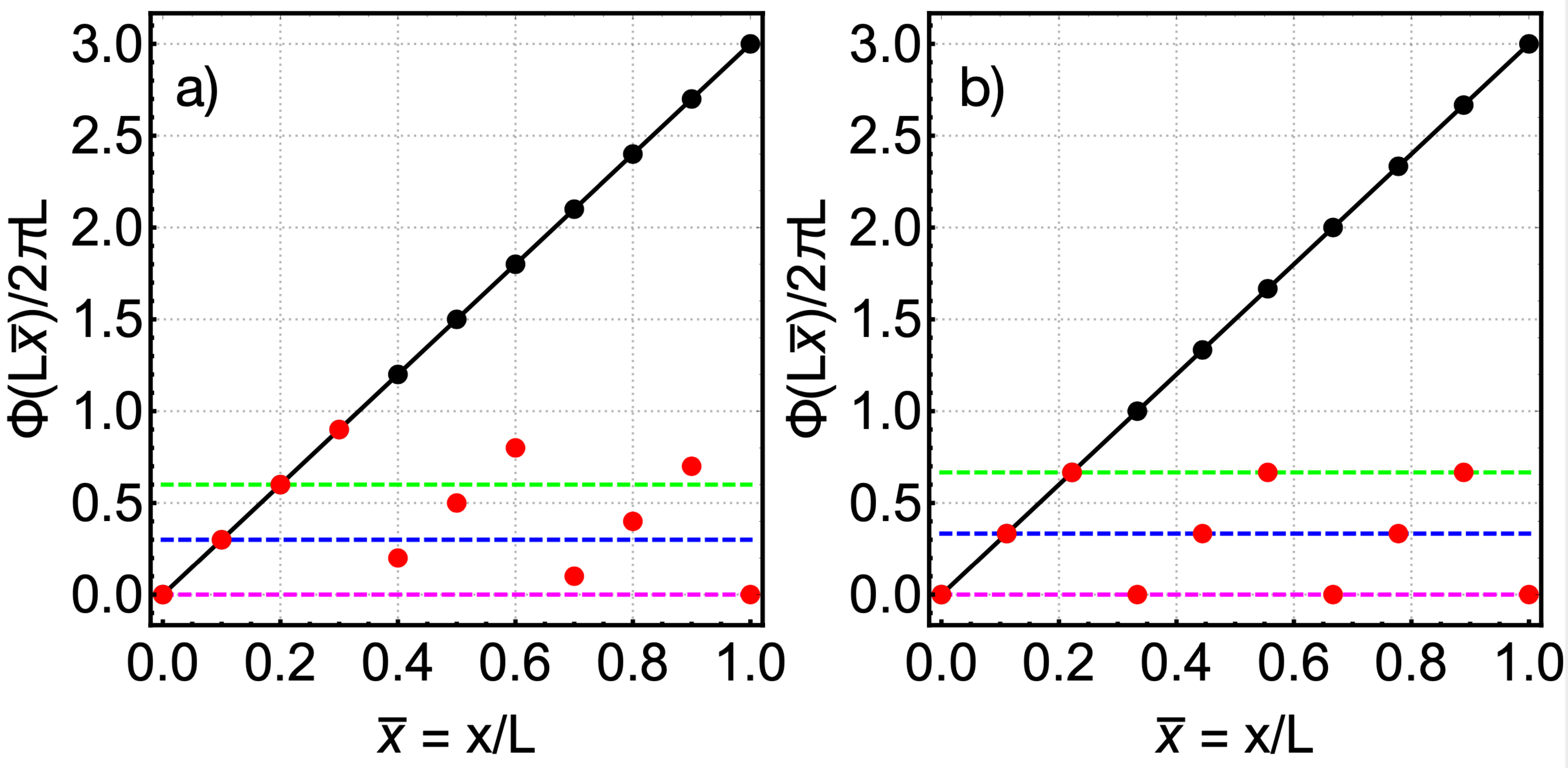}
   \end{center}
   \caption
   { \label{fig1b} a) Metasurface contribution to refraction. An incident plane-wave from a medium with refractive index $n_1$  is refracted to a medium with refractive index $n_2$. The metasurface is sketched by a gradient color from red to gray. The angle of incidence is $\theta_1$ whereas the refracted angle is $\theta_2$. The black and red lines represents respectively the wave-vector direction of the plane-waves in the medium of incidence and in the medium of transmission.   b) Metasurface made by a collection of diffractive elements (scatterers). Each dots represents a scatterer that radiates a plane wave with an extra phase-term $\Phi(x_j)$ as compared to the incident plane wave, $x_j$ being the scatterer position.}
\end{figure} 
 
We can now make a more accurate definition of the terms: phase profile $\Phi(x)$ and phase-jumps $\Phi(x_j)$. A metasurface is a photonic structure that implements a phase-jump $\Phi(x_j)$ to the scattered electric field at a position $x_j$ relatively to an incident plane-wave. The induced phase-jump varies locally from one point to another. The continuous function interpolating all the phase-jumps is the phase profile $\Phi(x)$. The scatterers realize the sampling of the phase profile $\Phi(x)$. This sampling can be made regular or random. The sampling leads to a sequence of phase jumps $\{ \Phi(x_j)\}_{j=0,...,N-1}$ where $\{x_j\}_{j = 0,...,N-1}$ are the positions of the scatterers that constitute the metasurface. 
\\
In case of a regular sampling, the position of the scatterers is periodic in space with period d, in such a way that two consecutive scatterers at position $x_j$ and $x_{j+1}$ respectively, are separated by a constant distance $d$, \textit{i.e.} $x_{j+1}-x_{j} = d$. However, this does not imply that the sequence of phase jumps $\{ \Phi(x_j) \}_{j=0,..,N-1}$ is periodic, in the sense that there exists $M \in \mathbb{N}$ such as $\Phi(x_j) = \Phi(x_{j+M})$ for $j \in \{0,\hdots,N-M-1\}$,  or equivalently $\Phi((j+M)d)=\Phi(j d)$. Assume then that $\Phi(x)$ is periodic with period $\Lambda$. Then the sequence $\{ \Phi(x_j) \}$ is periodic if and only if : $Md=n\Lambda$ for some $n \in \mathbb{N}$ and thus if and only if the periods $\Lambda$ and $d$ are such that $$\frac{d}{\Lambda} \in \mathbb{Q},$$ the period $M$ being defined as the smallest integer such that $M d/\Lambda \in \mathbb{N}$. This remark remains true even when the values of the phase $\Phi(x)$ are reduced modulo $2\pi$ to the interval $[0,2\pi]$. Let us denote this reduced phase by $\bar{\Phi}(x)$ and let us consider the case of a linear phase profile:  $\Phi(x)=\phi_1 x$. The reduced phase $\bar{\Phi}(x)$ is periodic with period $\Lambda$ defined by 
\begin{equation}
\phi_1 \Lambda = 2\pi \label{eq:LS1}.
\end{equation}
Indeed, the periodicity can be written: $\phi_1 (x + \Lambda) =  \phi_1 x + 2\pi$. Besides, the condition that the periods be in a rational ratio reads as
\begin{equation}
 k d/\Lambda \in \mathbb{N} , \hbox{ for some } k  \in \mathbb{N}  \label{eq:LS2}.
\end{equation}

If both conditions eq.(\ref{eq:LS1}) and eq.(\ref{eq:LS2}) are satisfied then the sampled linear phase profile behaves as a periodic sawtooth-wave profile.
This is illustrated in Fig.(\ref{fig1b}) where the linear phase profile $\Phi(x) = 3\times2\pi \frac{x}{L}$ is considered where $L$ is the total length of the metasurface. In that case $\Phi(x+\frac{L}{3})=\Phi(x)+2\pi$ and therefore $\bar{\Phi}(x)$ is periodic with period $\Lambda=L/3$.  The sampling is over the points $x_j=j d,\,j=0\hdots N-1$ and $d=L/(N-1)$. We obtain $d/\Lambda=3/(N-1)$. The rationality condition implies that, if $N-1$ is a multiple of $3$ then the sequence $\{ \Phi(x_j) \}$ has period $M=(N-1)/3$, otherwise the period is $N-1$, but this corresponds to the complete set of values $\{ \Phi(x_j) \}$ over the entire metasurface and effectively no periodicity is seen.
Indeed, as seen in Fig.(\ref{fig1b}-a), when the phase profile is sampled over $N=11$ scatterers, as shown by the black dots, the sequence of reduced phase-jumps $\{ \bar{\Phi}(x_j)\}_{j=0,..,N-1}$ is not periodic. On the contrary the same linear phase-profile, sampled over $N=10$ scatterers leads to a periodic sequence $\{ \Phi(x_j) \}$ with  $M=(10-1)/3=3$. In such a case, it can be described as a sawtooth-wave profile. This case is illustrated in Fig.(\ref{fig1b}-b). Because of the periodicity, sawtooth-wave profiles lead to drastic conditions for the definition of a generalized law of refraction. In particular, the generalized Snell-Descartes law found in this context is nothing else than the grating relation. This case is studied in details at the end of this paper. On the opposite, if the sequence of phase-jumps $\{ \Phi(x_j) \}_{j=0,..,N-1}$ does not exhibit any periodicity then the generalized Snell-Descartes law does not derive from the grating relation.

\begin{figure}
   \begin{center}
   \includegraphics[width=12cm]{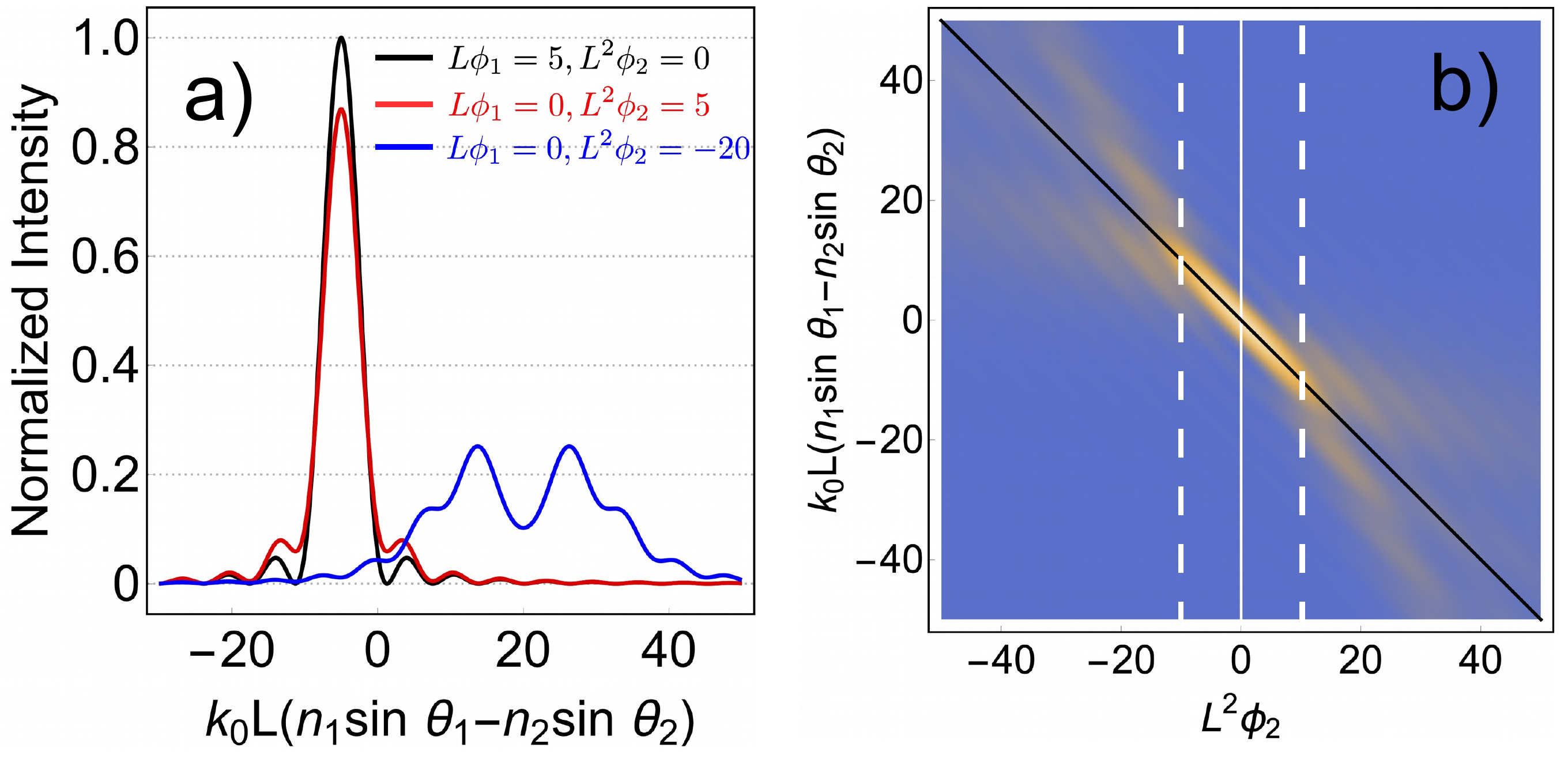}
   \end{center}
   \caption{\label{Fig:2} a) Light-intensity distribution as a function of $k_0 L D(\theta_1,\theta_2)$ for different phase-discontinuity profile: $\Phi(x) = 5 x $ ( black curve ), $\Phi(x) = 5 x^2 $ ( red curve ),$\Phi(x) = -20 x^2 $ ( blue curve ). b) Light-intensity distribution for parabolic phase-discontinuity function as a function of $k_0 L D(\theta_1,\theta_2)$ and $L^2\phi_2$ with $\phi_1=0$. The intensity scale is yellow for $I=1$ and blue for $I=0$.}
\end{figure} 

Let us now consider a parabolic phase-profile $\Phi(x)$ with ($\phi_k = 0 ~~\forall k>2$). In this case also an analytical formula for the scattered-field amplitude can be calculated [see eq.(S.11) in the supp. informations]. The light intensity scattered by a parabolic phase-profile is shown in Fig.(\ref{Fig:2}-a) for $\phi_2=5$ (red curve) and $\phi_2=-20$ (blue curve). For moderate values of $L^2 \phi_2  \lesssim 10 $, the light-intensity distribution exhibits only one single maximum. Its shape differs slightly from the linear case as shown by an inspection of the black curve ($L\phi_1=5, L^2\phi_2=0$) and the red curve ($L\phi_1=0, L^2\phi_2=5$) in Fig.(\ref{Fig:2}-a). The light intensity remains mainly localized near one maximum. The light scattered by the metasurface reduces to a single plane-wave, whose direction is given by a generalized Snell-Descartes law as shown analytically in the section III of Supp. Informations. [eq.(S.12)]:
\begin{align}
\label{eq:GenDesc2}
n_2 \sin \theta_2 - n_1 \sin \theta_1 & = \frac{1}{k_0} \phi_1 + \frac{L \phi _2}{k_0} .
\end{align}
This relationship holds for moderate values of the quadratic term $|L^2 \phi_2 | \leq 10$ as it can be seen in Fig.({\ref{Fig:2}-b) where the light-intensity distribution was calculated as a function of $k_0 L D(\theta_1,\theta_2)$ and $L^2 \phi_2$ with $\phi_1=0$. The dashed white lines enlights the region of validity of the eq.(\ref{eq:GenDesc2}) since the light intensity is located around one maximum leading to the radiation of a single plane-wave. In this range, the quantity $k_0 L D(\theta_1,\theta_2)$ equals the quantity $L^2 \phi_2$ as shown by the black line. Again, this generalized Snell-Descartes law does not agree with the previously reported one\cite{Aieta_2012b,Yu_2011}. 

In order to explain why a generalized law can only be approximatively defined for nonlinear phase profiles, we follow Jia \textit{et al.} \cite{Jia_2017}. The eq.(\ref{eq:ConSum}) can formally be integrated if the phase profile behaves as a power series. The result is given by:
\begin{align}
E(x,y) &= A_0 e^{i(\vec{k}_2.\vec{r}-\omega t)}  \lim_{y \to 0} e^{ i \sum_{k=1}^{\infty} \phi_kL^k\partial^k_y}~ \left ( \frac{e^{y+i D}-1}{y+i D} \right ) \nonumber \\
& = A_0 e^{i(\vec{k}_2.\vec{r}-\omega t)}  \lim_{y \to 0} \prod_{k=2}^\infty e^{ i \phi_k L^k \partial^k_y}~ \left ( \frac{e^{y+i D+i L \phi_1}-1}{y+i D+ i L \phi_1} \right ).
\end{align}

The linear term $\phi_1 L \partial_y$ acts as a shift operator [it is the generator of translations, \textit{i.e.} $e^{i\phi_1 L \partial_y} f(y) = f(y + i\phi_1 L)$ for any function $f(y)$], which explains why linear phase profiles shift the maximum of the light-intensity distribution by the quantity $L \phi_1$ and leads to a generalized Snell-Descartes law. Higher order terms $k \ge 2$ have a more complicated action that results in a shift and a distortion of the light-intensity distribution. As an example, $e^{i\phi_2 L^2 \partial^2_y}$ is the heat kernel with a time-like quantity $\tau \equiv i\phi_2 L^2$. It is known that through heat diffusion, the maximum of the heat distribution shifts but also that the amplitude decreases  as $\frac{1}{\sqrt{\tau}}$. For large values of the nonlinear term $\phi_2 L^2$, the light-intensity distribution is broad, as illustrated by the blue curve in Fig:(\ref{Fig:2}-a). In such a case,  we can no longer consider that a single plane-wave emerges from the metasurface. Thus only linear a phase $\Phi(x)$ rigorously leads to a generalization of the law of refraction, although it can be approximatively defined in the case of small nonlinear-terms.

A further inspection of the generalized law of refraction in the case of a linear profiles [eq.(\ref{eq:GenDesc})] explains why in the near-infrared and visible wavelengths range, periodic metasurfaces are needed in order to measure some effects \cite{Huang_2012,Li_2015,Ni_2011,Sun_2012,Yu_2011}. Indeed, the right-hand term is $\frac{\phi_1}{k_0}$. We denote $\Lambda$ the characteristic length-scale of the linear function $\Phi(x)$. It then reads $\Phi(L\bar{x}) = \alpha_1 \frac{L}{\Lambda} \bar{x}$ with $\phi_1 =  \frac{\alpha_1}{\Lambda}$. As a consequence, the right-hand term in eq.(\ref{eq:GenDesc}) reads $\frac{1}{k_0\Lambda} \alpha_1$. In order to get significant effects, one must have $k_0\Lambda \lesssim 10$ since $\alpha_1 \sim 1$. If the linear gradient extends over the whole metasurface length $\Lambda = L$ then it is only for large wavelengths, e.g. in the THz range or larger wavelengths, that such a metasurface will bend light in an anomalous direction as compared to the prediction of the Snell-Descartes law. For wavelengths smaller than the near-infrared, periodic metasurfaces are required because the characteristic length scale $\Lambda$ should be of the order of the wavelength of the incident light.

Our phenomenological model also applies to periodic metasurfaces. Let us consider a periodic phase-profile $\Phi(x)$  made of M scatterers per period and P periods. The total number of scatterers is then $N=P\times M$. We denote by $\tilde{\Phi}(x)$ the phase profile restricted to one period and $\Lambda$ the period. We assume that the scatterers are regularly spaced with distance $d$, in such a way that $\Lambda= M\times d$. This provides a periodic sampling of the phase profile, meaning that the scatterer at position $x_j+\Lambda$ induces the same phase-jump as the scatterer at position $x_j$, \textit{i.e.}  $\Phi(x_j+p \Lambda) = \Phi(x_j) = \tilde{\Phi}(x_j)$, $\forall x_j \in [0,\Lambda]$ with $p$ an integer in the range $p\in \{0,..,P-1\}$.Within this geometry, eq.(\ref{eq:DisSum}) reads:
\begin{align*}  
&E(x,y)= A_0 e^{i(\vec{k}_2.\vec{r}-\omega t)} \left(\sum_{p=0}^{P-1} e^{i k_0 d M D(\theta_1,\theta_2)p } \right)\times \left(\sum_{j=0}^{M-1} e^{i k_0d  D(\theta_1,\theta_2) j } ~ e^{ i \tilde{\Phi}(j d)} \right)
\end{align*}

Let us now assume that the phase function varies linearly over one period. The phase discontinuity is then a sawtooth function as depicted in Fig.(\ref{Fig:3}-a). The restriction of the phase discontinuity over one period is $\tilde{\Phi}(x)=\tilde{\phi}_1 x$ with $\tilde{\phi}_1 \in \mathbb{R}^*$. Within these assumptions, the previous relation now reads: 
\begin{align*} 
E(x,y) &\propto  A_0 e^{i(\vec{k}_2.\vec{r}-\omega t)}  S_P S_M\\
& S_P = \frac{\sin \left (\frac{P}{2} k_0\Lambda D(\theta_1,\theta_2) \right)}{\sin \left ( \frac{1}{2} k_0\Lambda D(\theta_1,\theta_2) \right)}  \\
& S_M =  \frac{\sin \left( \frac{1}{2} k_0\Lambda [D(\theta_1,\theta_2)+\frac{\tilde{\phi}_1}{k_0}] \right)}{\sin\left( \frac{1}{2} k_0 d [D(\theta_1,\theta_2)+\frac{\tilde{\phi}_1}{k_0}] \right)} .
\end{align*}

Some irrelevant phase factors have been removed from the previous equation. The contribution $S_P$ is a consequence of the periodicity of the metasurface. Its contribution to the light intensity is a comb-like curve [see blue curve in the insert in Fig.(\ref{Fig:3}-b)]. The contribution of the scatterers to the light intensity is the function $S_M$. This function is plotted as the red curve in the insert in Fig.(\ref{Fig:3}-b). Two consecutive maxima of the comb-like function $|S_P|^2$ are separated by $\Delta D(\theta_1,\theta_1) = \pm \frac{2\pi}{k_0 \Lambda}$. The two first-zeros on each side of the maximum of the function $|S_M|^2$ are separated by $\frac{2\pi}{k_0\Lambda}$. As a consequence only one plane wave emerges from the metasurface, provided that the following two conditions are satisfied:
\begin{align}
k_0 \Lambda D(\theta_1,\theta_2) = 2 m \pi \text{~with~} m \in \mathbb{Z}  \text{~and~} D(\theta_1,\theta_2)+\frac{\tilde{\phi}_1}{k_0} = 0 \nonumber
\end{align} 

Combining both equations, a generalized Snell-Descartes law can be defined if and only if $\tilde{\phi}_1$ takes the value $\tilde{\phi}_1=\tilde{\phi}_1^\star$ given by:
\begin{align}
\label{eq:Cond}
\tilde{\phi}_1^\star = \frac{2 m \pi}{ \Lambda} \text{~with~} |m| \le \lfloor{ \frac{k_0 \Lambda}{\pi} } \rfloor{},
\end{align} 
where $\lfloor x \rfloor$ is the integer-part of $x$. Within the previous conditions, the generalized Snell-Descartes law for periodic metasurfaces reads:
\begin{align}
\label{eq:GeneDesc3}
n_2 \sin \theta_2 - n_1 \sin \theta_1 & = \frac{\tilde{\phi}^\star_1}{k_0}  = \frac{\tilde{\Phi}(x)}{k_0 x}.
\end{align}
First, let us stress again that the generalized Snell-Descartes law [eq.(\ref{eq:GeneDesc3})] found in the context of periodic metasurfaces is not the eq.(\ref{law}) found in \cite{Aieta_2012b,Yu_2011} but most importantly it is in agreement with the experimental observations\cite{Huang_2012,Li_2015,Ni_2011,Sun_2012,Yu_2011}.  
Indeed, sawtooth profiles are piecewise linear and periodic. So the right-hand term $\frac{d \Phi(x)}{dx}$ in eq.(\ref{law}), found in \cite{Aieta_2012b,Yu_2011}, would imply Dirac distributions at each point where the phase profile is discontinuous. As a consequence for the periodic phase-profile $\Phi(x)$ with the restricted phase-profile $\tilde{\Phi}(x) = \tilde{\phi}_1 x$ defined previously, it would read  $\frac{d \Phi(x)}{dx} = \tilde{\phi}_1\left [1-\sum_{m=-\infty}^{+\infty} \delta(x-m\Lambda) \right ]$ where $\delta(x)$ is the Dirac distribution.
The contribution of the Dirac distributions was not measured in any experiment. On the opposite our expression eq.(\ref{eq:GeneDesc3}) is in agreement with the experimental results\cite{Ni_2011,Yu_2011,Huang_2012,Li_2015,Sun_2012} where periodic-metasurfaces add a contribution to the ray deviation that is proportional to $\tilde{\phi}_1^\star=\frac{2 \pi}{ \Lambda}$, the phase gradient restricted to one period augmented by the conditions eq.(\ref{eq:Cond}).

The conditions eq.(\ref{eq:Cond}) are drastic. Indeed, on the opposite to non-periodic metasurfaces with linear phase-profile (\textit{c.f.} eq.(\ref{eq:GenDesc})) the value of the slope $\tilde{\phi}_1$ cannot be chosen arbitrarily. The phase jump must vary between $0$ and  $2m\pi$ over one period $\Lambda$. 
Our phenomenological model explains why in all the experiments \cite{Huang_2012,Li_2015,Ni_2011,Sun_2012,Yu_2011} the function restricted to one period $\tilde{\Phi}(x)$ was chosen to vary between $0$ and $2\pi$~\cite{Li_2019}: this is the only way for a single plane-wave to emerge from the metasurface and, as a consequence, the only way to define a generalized law of refraction.  Finally, we emphasize that, combining eq.(\ref{eq:GeneDesc3}) and eq.(\ref{eq:Cond}), leads to the grating relation $D(\theta_1,\theta_2)=m \frac{2\pi}{k_0 \Lambda}$. Then, due to periodicity, the generalized law of refraction for periodic metasurfaces formally reduces to the grating relation. The equivalence between blazed diffraction gratings and metasurfaces was also established in the ref.\cite{Larouche_2012}, but in the continuum approximation only.

 \begin{figure}
   \begin{center}
   \includegraphics[height=8cm]{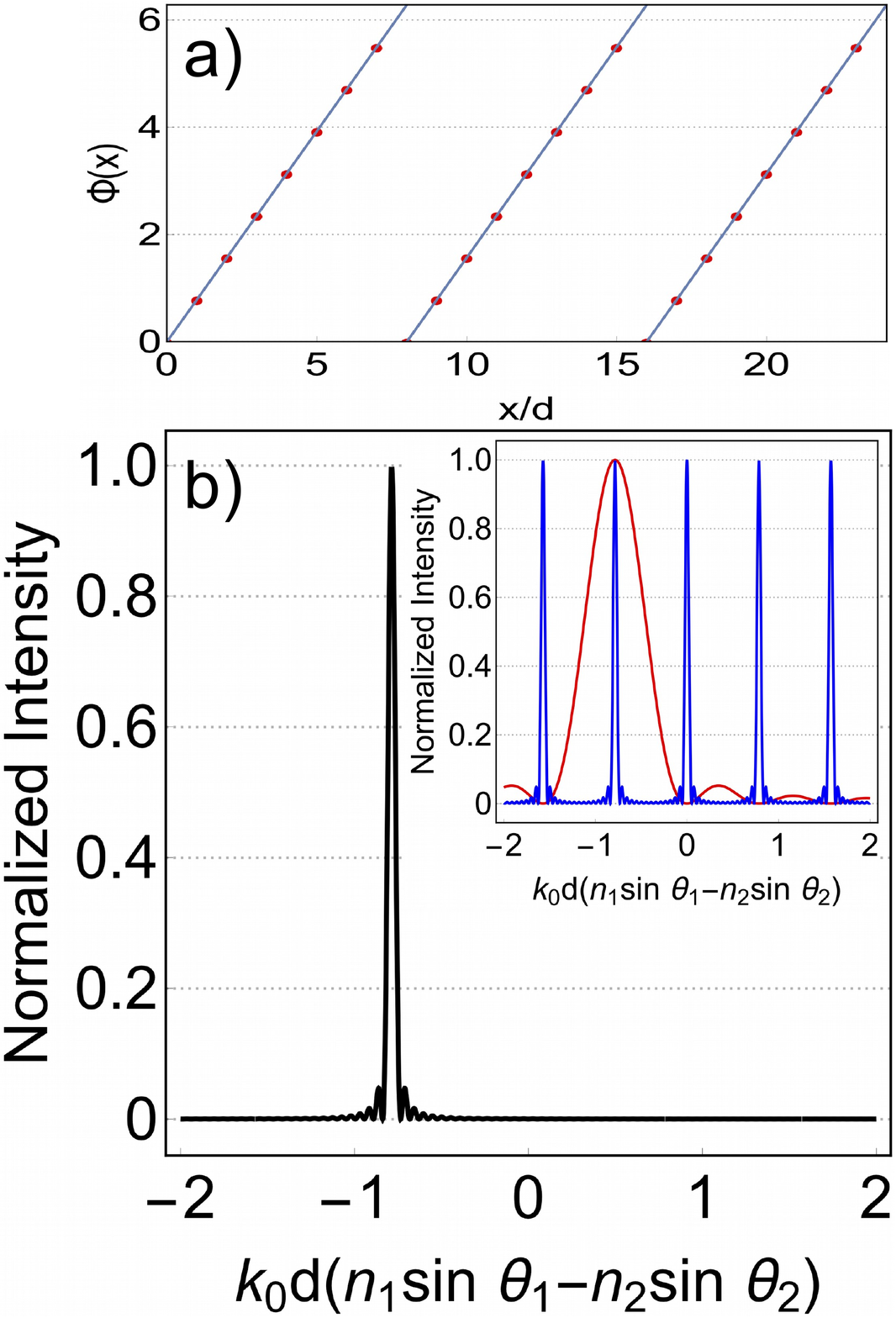}
   \end{center}
   \caption{\label{Fig:3} a) Sawtooth phase-discontinuity function for a periodic profile with $M=8$ scatterers (red dots) in one period (period length $\Lambda$) similar as the experimental realizations \cite{Ni_2011,Yu_2011}. b) Light-intensity distribution for linear phase-discontinuity function with $\tilde{\Phi}(x) = -2\pi \frac{x}{\Lambda}$. Insert: Contribution of the periodicity $|S_P|^2$ (Blue curve) and contribution of one period $|S_M|^2$ (red curve). }
\end{figure} 

\section{Conclusion}
We have derived three generalized law Snell-Descartes law as well as their domain of validity in the context of metasurfaces made of diffractive elements. They do not coincide with the previously reported results, except formally in the case of non-periodic metasurfaces with linear phase-profiles. Our results show that generalized laws of refraction are not unique and depend on the properties of the metasurface. Although simple to implement, our phenomenological model explains the main features of the available experiments. It could find applications in the design of metasurfaces beyond the search for generalized laws of refraction.

\part{\textit{The Supporting Informations}}
 \label{P.2}

\vspace{1cm}
\textbf{Abstract:}
This supplementary information provides some details on the calculations performed to obtain (i) the electric-field scattered by the metasurface, (ii) the generalized laws of refraction in the case of a linear phase-jump  $\Phi(x)=\phi_1 x$ and (iii) a parabolic phase-discontinuity $\Phi(x)=\phi_1 x + \phi_2 x^2$.
\vspace{1cm}

\setcounter{section}{0}
\section{Electric field scattered by the metasurface}

\begin{table}[h]
\begin{tabular}{| l | l |}
  \hline
symbol &~  \\
  \hline
 $\vec{r}$ = (x,y,0) & observation point  \\
 $k_0$ & magnitude of the wave-vector in vacuum  \\
 $\vec{k}_1= n_1 \vec{k}_0 $ & wave-vector in the medium with refractive index $n_1$ \\
 $\vec{k}_2= n_2 \vec{k}_0 $ & wave-vector in the medium with refractive index $n_2$ \\
 $\theta_1$ & angle of incidence \\
 $\theta_2$ & angle of refraction \\
 $D(\theta_1,\theta_2) = n_1 \sin \theta_1 - n_2 \sin \theta_2$ &  \\
 $(x_j,0)$ & position of the scatterer number $j$ \\
 $E_{inc}(x,y) = A_{inc} e^{i n_1 \vec{k}_0.\vec{r} }$ & incident plane-wave in the medium with homogeneous refractive index $n_1$ \\
 $E_j(x,y) = A_0 e^{i n_2 \vec{k}_0.\vec{r}} e^{i \Phi(x_j)}$ & plane-wave radiated by the scatterer number $j$ in the medium with homogeneous refractive index $n_2$\\
 $A_0$ & amplitude of the plane-wave scattered by the scatterers \\ 
 $N$ & number of scatterers \\
 $\Phi(x_j)$ & phase difference (or phase-jumps) added by the scatterer $j$ to the scattered plane-wave $E_j(x,y)$ \\
 & relatively to the incident plane-wave\\
 $\Phi(x)$ & phase profile, \textit{i.e.} a continuous function describing the phase-jumps induced by the scatterer\\
 $\{\Phi(x_j)\}$, $j=1,..,N$ & sampling of the phase function $\Phi(x)$ realized by the scatterers \\
 $\alpha_j \equiv \alpha$ & scattering efficiency\\
 $d$ & distance between two consecutive scatterers in case of a regular sampling\\
 $L$ & Total length of the metasurface \\
  \hline
 \end{tabular}
  \caption{List of symbol used in the article and in the supplementary materials.}
  \label{T1}
\end{table}

In this section we derive the electric field scattered by the metasurface that reads:
\begin{empheq}[box=\mybluebox]{align}
E(x,y,0)= A_0 e^{i(\vec{k}_2.\vec{r}-\omega t)} \sum_{j=0}^{N-1} e^{i k_0 D(\theta_1,\theta_2)x_j} e^{i \Phi(x_j)} \tag{S.1}
\label{eq:E}
\end{empheq}
~\\
The quantities are defined in the table  (\ref{T1}).
~\\

~\\
~\\
The metasurface is assumed to be at position $y=0$. The angle of incidence and refraction are shown on Fig.(1) in the main document. We assume that the first scatterer is at position $(x_0=0,0)$ that is chosen to be the origin of the phase.

We work in the scalar approximation and do not take into account any polarization effects. The model assumes that the particles are punctual (null size) and that their scattering efficiency $\alpha$ is identical. The metasurface is illuminated by a plane-wave in the medium with refractive index $n_1$, with an incident angle $\theta_1$ measured from the normal to the metasurface and an amplitude $A_{inc}$. The incident plane-wave at position $(x_j,0,0)$ reads:
\begin{align*} 
E_{inc}(x_j,0,0)= A_{inc} e^{i(n_1 k_0 \sin \theta_1 x_j-\omega t)} 
\end{align*}
It drives the electric charges in the particle number $j$ that radiates an electric-field that is assumed to be a spherical wave centered on the particle position $(x_j,0,0)$. The field scattered by the particle $j$ at the observation point $(x,y,z)$ reads:
\begin{align*} 
E_{j}(x,y,z)= \alpha A_{inc} e^{i(n_1 k_0 \sin \theta_1 x_j-\omega t)} \frac{1}{n_2 k_0\sqrt{(x-x_j)^2+y^2+z^2}} e^{i n_2 k_0 \sqrt{(x-x_j)^2+y^2+z^2}} e^{i \Phi(x_j)}
\end{align*}

We have assumed that the field radiated by the scatterer is delayed as compared to the incident plane-wave\cite{Genevet_2017}. This results in an intrinsic phase $\Phi(x_j)$ that depends on the scatterer. Engineering the scatterers geometry leads to a sampling of the phase function $\Phi(x)$ leading to a list $\{ \Phi(x_j), j=1,...,N\}$.

We further assume that the metasurface is invariant in the $z$-direction. We then consider only the plane $z=0$. In this plane, we denote by $\vec{k}_2 = n_2 k_0 \sin \theta_2 \vec{e}_x + n_2 k_0 \cos \theta_2 \vec{e}_y$ a wave-vector in the medium with refractive index $n_2$ making the angle $\theta_2$ with the normal $\vec{e}_y$ to the metasurface. We consider the observation point M with coordinates $(x,y)$, \textit{i.e.} $\vec{OM} = \vec{r}=x \vec{e}_x + y \vec{e}_y$. The scatterer is at position $\vec{OP}= x_j \vec{e}_x$. We denote $\vec{R} = \vec{OM} - \vec{OP} = (x-x_j)\vec{e}_x + y \vec{e}_y$, the vector joining the scatterer and the observation point. We call $\varphi$ the angle between the vectors $\vec{R}$ and $\vec{k}_2$. The previous quantities are defined on the Fig.(\ref{Fig:Sph_To_Pl}). We can perform the following approximation $n_2 k_0 |\vec{R}| \simeq \vec{k}_2.\vec{R}$ if the angle $\varphi \ll 1$. Indeed, $ \vec{k}_2.\vec{R} = n_2 k_0 |\vec{R}| \cos \varphi = n_2 k_0 |\vec{R}| + o(\varphi)$. This approximation is valid in first order in $\varphi$. Then we have to assume the paraxial approximation that leads to considering only observation points close to the normal $\vec{e}_y$ such as $\left |\frac{x_j}{y} \right | \ll 1$. This corresponds to the conditions of the Fraunhofer diffraction regime\cite[p.382]{Born_1985} and ref.\cite[p.465]{Hecht_2015}. 

\begin{figure}[htbp]
\begin{center}
\includegraphics[width=5cm]{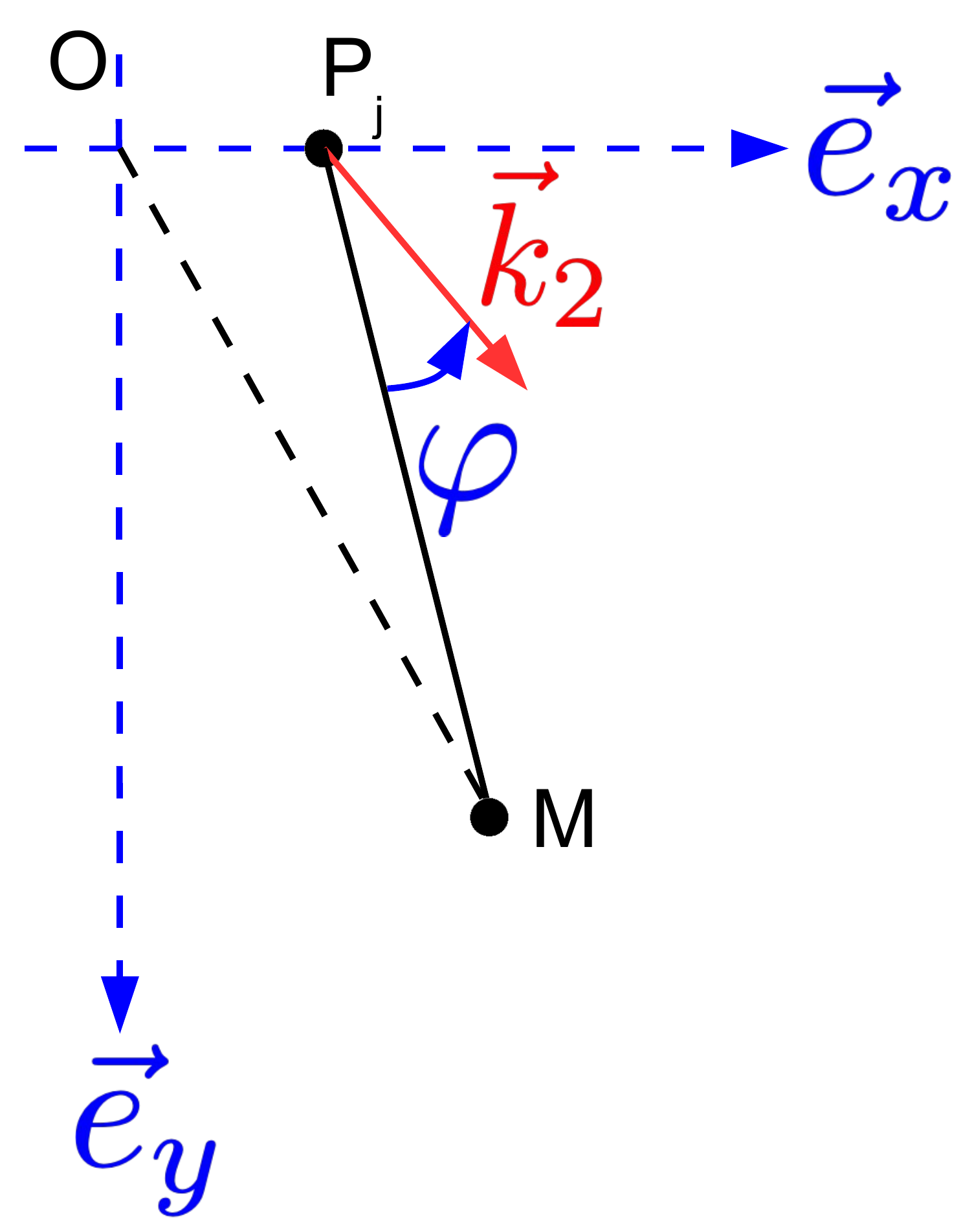}
\caption{Notations sustaining the approximations from the spherical-wave to the plane-wave. The scatterer is at the position $P_j$, the observation point is M=(x,y). We consider a wave-vector $\vec{k}_2$ making an angle $\theta_2$ with the axis $\vec{e}_y$. The angle between the vector $\vec{k}_2$ and the vector $\vec{PM}$ is $\varphi$.}
\label{Fig:Sph_To_Pl}
\end{center}
\end{figure}

Within the preceding approximations the electric-field scattered by the particle number $j$ reduces to a plane-wave that reads:
\begin{align*} 
&E_{j}(x,y,0) \simeq A_{0} e^{i(n_1 k_0 \sin \theta_1 x_j-\omega t)}  e^{i n_2 k_0 [(x-x_j) \sin \theta_2+y \cos \theta_2]} e^{i \Phi(x_j)} \\
&E_{j}(x,y,0) \simeq A_{0} e^{i [ k_0 (n_1 \sin \theta_1 - n_2 \sin \theta_2 )x_j-\omega t]}  e^{i n_2 k_0 [x \sin \theta_2 + y \cos \theta_2]} e^{i \Phi(x_j)} \\
&E_{j}(x,y,0) \simeq A_{0} e^{i \vec{k}_2.\vec{r}-i \omega t}  e^{i  k_0 D(\theta_1,\theta_2)x_j} e^{i \Phi(x_j)} 
\end{align*}

with $D(\theta_1,\theta_2) = n_1 \sin \theta_1 - n_2 \sin \theta_2$ and $A_0 = \alpha A_{inc} \frac{1}{n_2 k_0 |\vec{R}|} \simeq  \alpha A_{inc} \frac{1}{n_2 k_0 y}$ the amplitude of the plane-wave that is constant in a plane of constant $y$.

We now sum the contribution of the N scatterers $E(x,y,0) = \sum_{j=0}^{N-1} E_j(x,y,0)$, which leads to the eq.(\ref{eq:E}).

Notice that our model consists in summing the contribution of N coherent oscillators in the spirit of the ref.\cite[chap.30]{Feynman_2011} and the ref. \cite[p.462]{Hecht_2015} but with an intrinsic phase-shift $\Phi(x_j)$ that is dependent on the scatterer position $x_j$. 

\section{Linear phase profile}

\subsection{Regular Sampling}

\subsubsection{Light-intensity distribution}

We consider an array of N equally-spaced scatterers separated by a distance $d$. The position of the scatterer number $j$ on the metasurface abscissa is $x_j = j d$ with $j \in \{0,...,N-1\}$. We assume a linear phase-function $\Phi(x) = \phi_1 x$. Following these assumptions, the discrete sum in eq.(\ref{eq:E}) can be performed. The light intensity reads:

\begin{align}
\left | \frac{E(\theta_2)}{A_0} \right |^2 &= \frac{\sin^2 \left [\frac{1}{2}N k_0 d \right (D(\theta_1,\theta_2) + \frac{\phi_1}{k_0} \left ) \right]}{\sin^2 \left [\frac{1}{2}   k_0 d \right (D(\theta_1,\theta_2) + \frac{\phi_1}{k_0} \left ) \right]} \tag{S.2}
\label{eq:Id} 
\end{align}

The light-intensity distribution depicted in Fig.(\ref{Fig:Disc_Sum_Int}) is a comb-like function with subsidiary maxima between the principal peaks.  

\begin{figure}[htbp]
\begin{center}
\includegraphics[width=8cm]{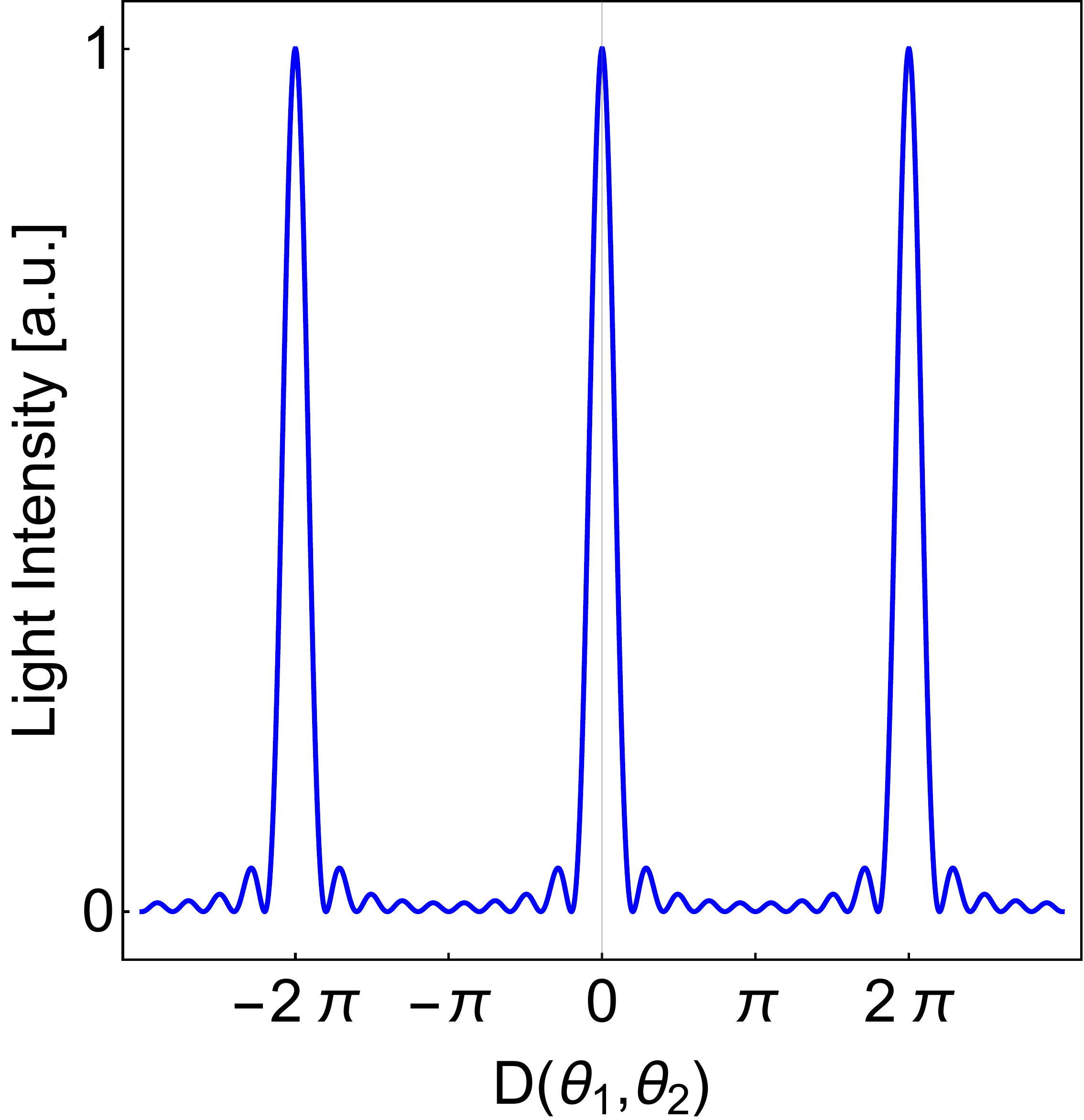}
\caption{Light-intensity distribution eq.(\ref{eq:Id}) for N=10 scatterers. It has been normalized by $1/N^2$.}
\label{Fig:Disc_Sum_Int}
\end{center}
\end{figure}

\subsubsection{Conditions for a generalized Snell-Descartes law.}

In the regular sampling case, the light-intensity scattered by the metasurface reaches a maximum when the denominator of the eq.(\ref{eq:Id}) is null, \textit{i.e.} when the following relationship holds:
\begin{align}
\label{eq:GenDesc}
&n_2 \sin \theta_2 - n_1 \sin \theta_1 = \frac{1}{k_0} \phi_1 - m \frac{2\pi}{k_0 d} \tag{S.3} \\
&\text{with}\hspace{2mm} m \in \mathbb{Z} \nonumber
\end{align}

A generalized law of refraction can be defined if $\forall \theta_1 \in [0,\frac{\pi}{2}]$ only the term $m=0$ leads to a single refracted angle $\theta_2$, \textit{i.e.} only one refracted angle satisfies $\sin \theta_2 \in [-1,1]$. In such a case, the generalized law of refraction reads:
\begin{empheq}[box=\mybluebox]{align}
\label{eq:GenDesc2}
&n_2 \sin \theta_2 - n_1 \sin \theta_1 = \frac{1}{k_0} \phi_1 \tag{S.4}
\end{empheq}

In order to determine the conditions of existence of the generalized law of refraction, we define the quantity $B_m =  \frac{1}{k_0} \phi_1 - m \frac{2\pi}{k_0 d}$ and plot $\sin \theta_2 = f(\sin \theta_1)$. This is a linear function with slope $n_1/n_2 >0$ and vertical-intercept $B_m$.
\begin{align*}
\sin \theta_2 = \frac{n_1}{n_2} \sin \theta_1 + B_m
\end{align*}

A single plane-wave emerges from the metasurface if $B_{-1}>1$ and $\frac{n_1}{n_2} + B_{+1} < -1$ as shown by the Fig.(\ref{Fig:Desc_Grat}).

\begin{figure}[htbp]
\begin{center}
\includegraphics[width=8cm]{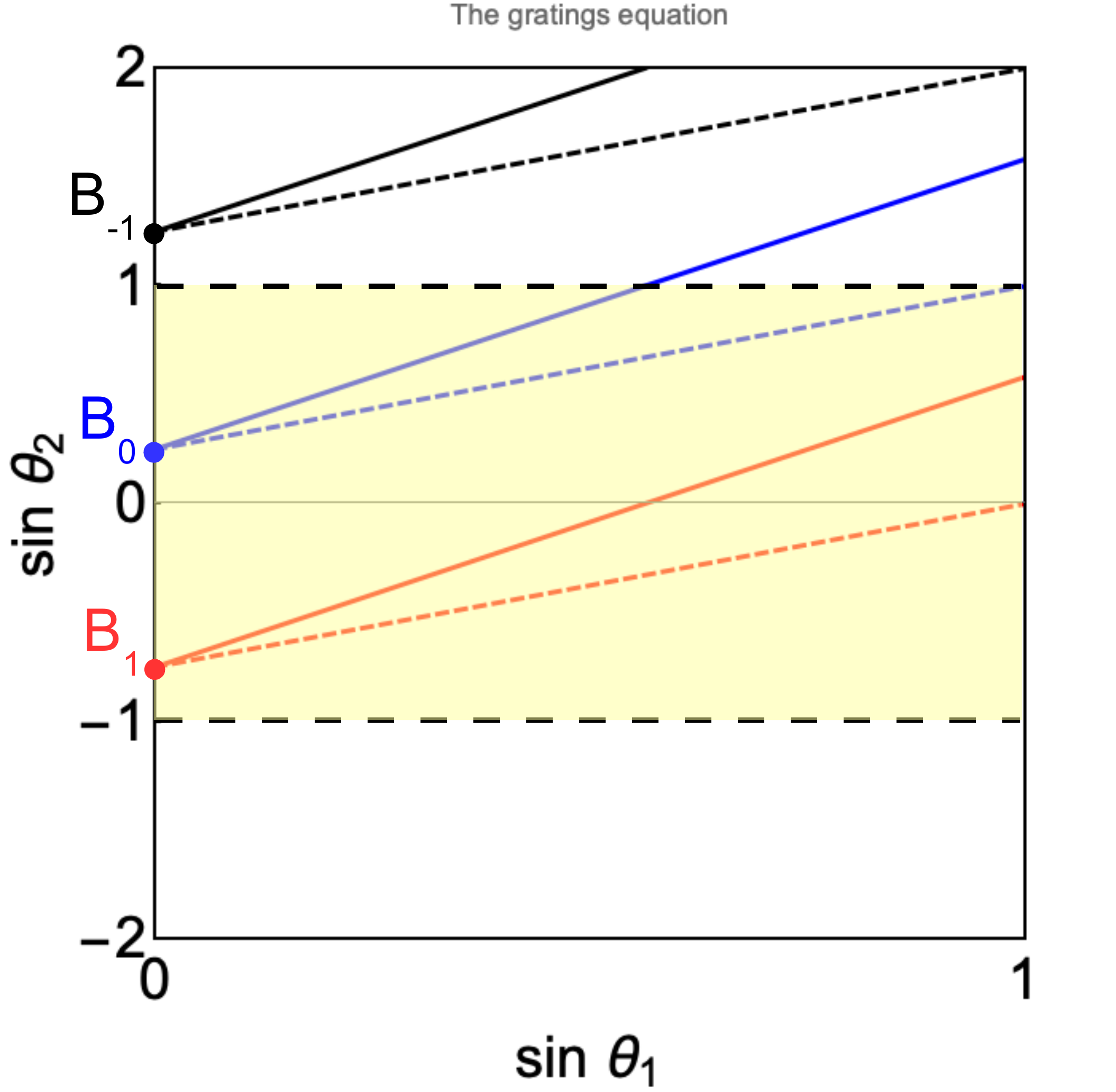}
\caption{diffraction order $m=\{ -1,0,1\}$. Plain lines $n_1 > n_2$, dashed lines $n_1 < n_2$. We assume that $\phi_1>0$.}
\label{Fig:Desc_Grat}
\end{center}
\end{figure}

The previous inequalities lead to:
 \begin{align*}
 \frac{1}{k_0} \phi_1 + \frac{2\pi}{k_2 d} >1 \hspace{2mm} \text{and}  \hspace{2mm}  \frac{n_1}{n_2}+\frac{1}{k_0} \phi_1 - \frac{2\pi}{k_2 d} < -1
\end{align*}
They restrict the range of admissible values for the slope of the linear phase-shift $\phi_1$ and the inter-particles spacing $d$ to 
 \begin{align}
&n_2 k_0 - \frac{2\pi}{d} < \phi_1 < \frac{2\pi}{d} - (n_1 +n_2)k_0 \hspace{2mm} \text{and} \tag{S.5}  \label{eq:Cd1} \\
& 0< k_0 d < \frac{2\pi}{n_2 + \frac{n_1}{2}} \tag{S.6} \label{eq:Cd2}
\end{align}
Note that $\phi_1$ can take large values in $]-\infty,+\infty[$ in the limit where $d \to 0$. As a consequence, these conditions do not hold in the continuum approximation.

In order to consider that only one plane-wave emerges from the metasurface the previous conditions are necessary but not sufficient conditions . Indeed, the principal maxima in the light-intensity distribution Fig.(\ref{Fig:Disc_Sum_Int}) have an angular width. For a fixed incident angle $\theta_1$, the angles corresponding to the first zeros on each side of the principal maximum with $m=0$ are:
\begin{align*}
n_2 \sin \theta_2^+ = n_1 \sin \theta_1 +  \frac{1}{k_0} \phi_1 + \frac{2\pi}{N k_0 d} \\
n_2 \sin \theta_2^- = n_1 \sin \theta_1 +  \frac{1}{k_0} \phi_1 - \frac{2\pi}{N k_0 d}
\end{align*}

These two angles can be considered to be equal if:
\begin{align*}
n_2 \sin \theta_2^+ - n_2 \sin \theta_2^- = \frac{4\pi}{N k_0 d} \ll 1
\end{align*}

Differencing two plane-waves is a property of the measuring device (the observer). Denoting by $\epsilon \ll 1$ the angular resolution of the observer, the minimum number of scatterers must be:
\begin{align}
N \ge \frac{4\pi}{\epsilon k_0 d} \tag{S.7} \label{eq:Cd3}
\end{align}

in order to consider that a single plane-wave emerges from the metasurface. As a consequence only in the limit $N \to +\infty$, a generalized Snell-Descartes law can be defined  independently of the observer's properties. 

\subsection{The continuum approximation}

For a random  as well as a regular sampling but in the continuum approximation, the light-intensity distribution can be computed from the eq.(3) in the main document. It reads:
\begin{align}
\left | \frac{E(\theta_2)}{\tilde{A}_0} \right|^2 &= \text{sinc}^2 \bigg [\frac{1}{2} k_0 L \big (D\left(\theta_1,\theta_2\right) + \frac{\phi_1}{k_0} \big ) \bigg] \tag{S.8} \label{eq:Ic}
\end{align}

$\text{sinc}$ being the sinc-function. The intensity distribution is characterized by a principal maximum for a null argument of the sinc-function. It leads to a generalized law of refraction:

\begin{empheq}[box=\mybluebox]{align}
\label{eq:GenDesc2}
&n_2 \sin \theta_2 - n_1 \sin \theta_1 = \frac{1}{k_0} \phi_1 \tag{S.9}
\end{empheq}

Again considering that a single plane-wave emerges from the metasurface depends on the angular resolution of the observer. In the continuum approximation this criteria leads to a minimal length of the metasurface:
\begin{align}
k_0 L \ge \frac{4\pi}{\epsilon} \tag{S.10} \label{eq:Cc1}
\end{align}

As a consequence only in the limit $k_0L \to +\infty$, the generalized Snell-Descartes law can be defined  independently of the observer's properties.

To conclude, a generalized law of refraction can be defined in the case of a phase-shift that varies linearly along the metasurface. It takes the same form in the discrete case and in the continuum approximation. Nevertheless, some conditions prevail to its existence: the equations (\ref{eq:Cd1}-\ref{eq:Cd3}) in the discrete case and the equation  (\ref{eq:Cc1}) in the continuum approximation.

\section{Parabolic phase-shift profile: Light intensity distribution}

In this section we derive the transmitted electric-field in the case of a parabolic phase-shift profile where the phase function $\Phi(x)$ is assumed to be quadratic, \textit{i.e.} $\phi(x) = \phi_1 x + \phi_2 x^2$ with $\phi_2 \neq 0$. We then simplify the analytical expression in the limit where $\phi_2 \to 0$. We are finally able to derive the light-intensity distribution and to deduce a generalized Snell-Descartes law for refraction in the case of a parabolic phase-profile.

The electric-field  can be analytically computed in the case of the continuum approximation. It reads:

\begin{align}
&E(x,y) =   \tilde{A}_0 e^{i(\vec{k}_2.\vec{r}-\omega t)} \int_{0}^1 e^{i k_0 L D(\theta_1,\theta_2)\bar{x}} e^{i ( i  \phi_1 L \bar{x} + \phi_2 L^2 \bar{x}^2)} d\bar{x} \nonumber \\
&E(x,y) = \tilde{A}_0 e^{i(\vec{k}_2.\vec{r}-\omega t)} ~ \frac{e^{\frac{3i\pi}{4}} }{L \sqrt{\phi _2}}  \left [ F\left(\frac{e^{\frac{i\pi}{4}} \left(k_0 D(\theta_1,\theta_2)+\phi _1\right)}{2 \sqrt{\phi _2}}\right)-e^{i L \left[ k_0 D(\theta_1,\theta_2) + \phi _1 + L \phi _2 \right]} F\left(\frac{e^{\frac{i\pi}{4}} \left(k_0 D(\theta_1,\theta_2)+\phi _1 + 2 L \phi_2 \right)}{2 \sqrt{\phi _2}}\right) \right ] \tag{S.11}
\label{eq:E_Para}
\end{align}

Where $F(x)=e^{-x^2} \int_0^x e^{t^2} dt$ is the Dawson function. It has the property that $F'(x)=1-2x F(x)$. 

We now assume that $2 L \phi_2 \ll k_0 D(\theta_1,\theta_2) + \phi_1$. As a consequence, defining $Q = \frac{e^{\frac{i\pi}{4}} \left(k_0 D(\theta_1,\theta_2)+\phi _1 \right)}{2 \sqrt{\phi _2}}$
\begin{align*}
F(Q + e^{\frac{i\pi}{4}} L \sqrt{\phi _2} ) \simeq F( Q ) + [1-2 Q F(Q)] e^{\frac{i\pi}{4}} L \sqrt{\phi_2}
\end{align*}

We can now simplify the expression of the electric field in the limit where $\phi_2 \to 0$. It reads:

\begin{align*}
E(x,y) =& \tilde{A}_0 e^{i(\vec{k}_2.\vec{r}-\omega t)} ~ \frac{e^{\frac{3i\pi}{4}} }{L \sqrt{\phi _2}}  \left [ F(Q)-e^{i L \left[ k_0 D(\theta_1,\theta_2) + \phi _1 + L \phi _2 \right]} F(Q +  e^{\frac{i\pi}{4}} L \sqrt{\phi _2} ) \right ] \\
E(x,y) \simeq & \tilde{A}_0 e^{i(\vec{k}_2.\vec{r}-\omega t)} ~ \frac{e^{\frac{3i\pi}{4}} }{L \sqrt{\phi _2}}  \left \{ F(Q)-e^{i L \left[ k_0 D(\theta_1,\theta_2) + \phi _1 + L \phi _2 \right]} \left[ F( Q ) + (1-2 Q F(Q) ) e^{ \frac{i\pi}{4} } L \sqrt{\phi_2} \right] \right \} \\
E(x,y) \simeq& \tilde{A}_0 e^{i(\vec{k}_2.\vec{r}-\omega t)} ~ \frac{e^{\frac{3i\pi}{4}} }{L \sqrt{\phi _2}} F(Q)  \left \{1 - e^{i L \left[ k_0 D(\theta_1,\theta_2) + \phi _1 + L \phi _2 \right]}  \right \} - \tilde{A}_0 e^{i(\vec{k}_2.\vec{r}-\omega t)} ~ e^{\frac{3i\pi}{4}} e^{i L \left[ k_0 D(\theta_1,\theta_2) + \phi _1 + L \phi _2 \right]} \left[1-2 Q F(Q) \right] e^{ \frac{i\pi}{4} }  \\
E(x,y) \simeq& \tilde{A}_0 e^{i(\vec{k}_2.\vec{r}-\omega t)} ~ \frac{e^{\frac{3i\pi}{4}} }{L \sqrt{\phi _2}} F(Q)  \left \{1 - e^{i L \left[ k_0 D(\theta_1,\theta_2) + \phi _1 + L \phi _2 \right]}  \right \}
\end{align*}

In the last line we kept only the dominant term in $\phi_2$. Within this approximation, the light-intensity distribution reads:
\begin{align*}
\left | \frac{E(\theta_2)}{\tilde{A}_0 } \right |^2 \simeq  \frac{4}{L^2 \phi_2} |F(Q)|^2 \sin^2 \left[\frac{k_0 L}{2} \left(  D(\theta_1,\theta_2) + \frac{\phi _1}{k_0} + \frac{L \phi _2}{k_0} \right)\right]
\end{align*}

The light-intensity distribution is maximum for:
\begin{empheq}[box=\mybluebox]{align}
\label{eq:GenDesc2}
&n_2 \sin \theta_2 - n_1 \sin \theta_1 = \frac{1}{k_0} \phi_1 + \frac{L \phi _2}{k_0} \tag{S.12}
\end{empheq}
 
This is a generalized Snell-Descartes for refraction in the case of a parabolic phase-shift profile. This equation is numerically checked in the main letter as the Fig.(2b). It holds for $|L^2 \phi_2| \le 10$.
 
Since for small arguments, $F(Q) \simeq Q$, the intensity at maximum reaches the value:
 
\begin{align*}
\left | \frac{E(\theta_2)}{\tilde{A}_0 } \right |^2 \simeq  \frac{4}{L^2 \phi_2} \left | \frac{L \sqrt{\phi_2}}{2} \right |^2 = 1
\end{align*}

\bibliographystyle{acm}
\bibliography{HAL_Meta_Descartes}
 
\end{document}